\documentclass[letterpaper,twocolumn,prl,aps,showpacs,superscriptaddress,floatfix,nofootinbib]{revtex4-1}
\usepackage[latin1]{inputenc}
\usepackage{bm}
\usepackage{multirow}
\usepackage{amssymb}
\usepackage{amsbsy}
\usepackage{amsmath}
\usepackage{stmaryrd}
\usepackage{graphicx}
\usepackage{epsfig}
\usepackage{placeins}
\usepackage{ulem}
\usepackage[usenames,dvipsnames]{color}
\usepackage[colorlinks,linkcolor=blue,citecolor=blue,urlcolor=blue]{hyperref}
\makeatletter

\begin{document}
\title{Typicality Approach to the Optical Conductivity in \\  Thermal and Many-Body Localized Phases}

\author{Robin Steinigeweg}
\email{rsteinig@uos.de}
\affiliation{Department of Physics, University of Osnabr\"uck, D-49069 Osnabr\"uck, Germany}

\author{Jacek Herbrych}
\email{jacek@physics.uoc.gr}
\affiliation{Department of Physics, University of Crete, GR-71003 Heraklion, Greece}
\affiliation{Cretan Center for Quantum Complexity and Nanotechnology, University of Crete, GR-71003 Heraklion, Greece}

\author{Frank Pollmann}
\email{frankp@pks.mpg.de}
\affiliation{Max-Planck-Institut f\"ur Physik komplexer Systeme, D-01187 Dresden, Germany}

\author{Wolfram Brenig}
\email{w.brenig@tu-bs.de}
\affiliation{Institute for Theoretical Physics, Technical University Braunschweig, D-38106 Braunschweig, Germany}

\date{\today}


\begin{abstract}
We study the frequency dependence of the optical conductivity $\text{Re} \,
\sigma(\omega)$ of the Heisenberg spin-$1/2$ chain in the thermal and near
the transition to the many-body localized phase induced by the strength of a 
random $z$-directed magnetic field. Using the method of dynamical quantum
typicality, we calculate the real-time dynamics of the spin-current
autocorrelation function and obtain the Fourier transform $\text{Re} \,
\sigma(\omega)$  for system sizes much larger than accessible to
standard exact-diagonalization approaches. We find that the low-frequency
behavior of $\text{Re} \, \sigma(\omega)$ is well described by $\text{Re} \,
\sigma(\omega) \approx \sigma_\text{dc} + a \, |\omega|^\alpha$, with $\alpha
\approx 1$ in a wide range within the thermal phase and close to the transition. 
We particularly detail the decrease of $\sigma_\text{dc}$ in the thermal phase as
a function of increasing disorder for strong exchange anisotropies. We further
find that the temperature dependence of $\sigma_\text{dc}$ is consistent with
the existence of a mobility edge.
\end{abstract}
\maketitle

{\it Introduction.}
Many-body localization (MBL) generalizes the concept of Anderson localization
\cite{anderson1958} to interacting systems. In a pioneering work \cite{basko2006},
Basko, Aleiner, and Altshuler showed perturbatively that the Anderson insulator
is stable to small interactions. Thus, an isolated quantum many-body system
can undergo a dynamical phase transition from a thermal phase to an MBL phase
where eigenstate thermalization \cite{deutsch1991, srednicki1994, rigol2008}
breaks down. Subsequent numerical works further revealed the richness
of disordered many-body systems \cite{oganesyan2007, pal2010, bera2015, luitz2015}.
A characteristic property of MBL systems is a logarithmic growth of entanglement
after a global quench \cite{znidaric2008, bardarson2012}, which has lead to a
phenomenological understanding in terms of locally conserved quantities \cite{vosk2013,
serbyn2013, huse2014}. An exciting aspect of MBL is that it allows to protect quantum
orders at finite energy densities (both symmetry breaking and topological ones), which
would melt in thermal phases \cite{huse2013, kjall2014, pekker2014, chandran2014,
bahri2015}. On the experimental side, first observations of MBL in optical-lattice
systems have been made by studying quantum quenches in disordered systems of
interacting particles \cite{schreiber2015}. Furthermore, the $I$-$V$
characteristics of amorphous iridium-oxide reveal an insulating state where
MBL might play a role \cite{ovadia2015}.

\begin{figure}[t]
\centering
\includegraphics[width=0.90\columnwidth]{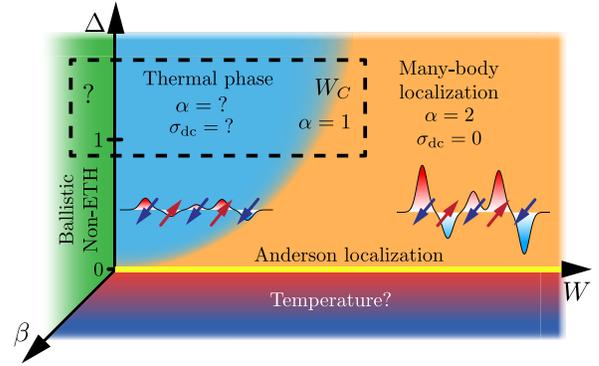}
\caption{(Color online) Dynamical phase diagram (sketch) of disordered
spin-$1/2$ XXZ chains. Issues studied in this paper: Scaling of dc
conductivity $\sigma_\text{dc}$ and low-frequency exponent $\alpha$
for strong interactions $\Delta \geq 1$ and disorders $W \geq 0$ up
to the MBL transition; temperature dependence and existence of mobility
edge; typicality in finite systems with $W > 0$.}
\label{Fig1}
\end{figure}

In the ongoing discussion of MBL, a central model is the spin-$1/2$ XXZ chain
with a spatially random $z$-directed magnetic field, being equivalent to interacting
spinless fermions in a random on-site potential of strength $W$. Furthermore, the XXZ
chain is a fundamental model for the study of transport and relaxation in low dimensions
\cite{review} and relevant to the physics of quasi one-dimensional quantum magnets
\cite{johnston2000, sologubenko2000, hess2001, hess2007, hlubek2010, thurber2001},
cold atoms in optical lattices \cite{trotzky2008}, and nanostructures
\cite{gambardella2006}, as well as to physical questions in a much
broader context \cite{kruczenski2004, kim1996}. This model is also of
paramount interest due to its remarkably rich dynamical phase diagram,
manifest in the frequency- and temperature-dependent optical conductivity
$\sigma(\omega, T)$. Despite integrability of the disorder-free XXZ chain,
$W = 0$, the exact calculation of $\sigma(\omega, T)$ at $T \neq 0$ has been
and continues to be a challenge to theory and is an important goal of new
analytical and numerical techniques. While it has become clear that, for small
particle-particle interactions $\Delta < 1$, $\sigma(\omega,T)$ features a
non-dissipative Drude contribution at $\omega = 0$ and any $T \geq 0$
\cite{shastry1990, narozhny1998, zotos1999, benz2005, fujimoto2003, prosen2011,
prosen2013, herbrych2011, karrasch2012, karrasch2013-1, steinigeweg2014-1,
steinigeweg2015-1, carmelo2015}, much less is known on the dynamics at $\omega
\neq 0$. Yet, signatures of diffusion, e.g., with a well-behaving limit $\omega
\to 0$, have been found only for strong $\Delta > 1$ and high $T$
\cite{znidaric2011, steinigeweg2011, karrasch2014} as well as for $\Delta = 1$
and very low $T$ \cite{sirker2009, sirker2011, grossjohann2010}.

Perturbations, such as spin-phonon coupling \cite{shimshoni2003, rozhkov2005,
hlubek2012}, dimerization \cite{huang2013, karrasch2013-2}, interactions between
further neighbors \cite{heidrichmeisner2003, steinigeweg2013} or different
chains \cite{sologubenko2000, hess2001, zotos2004, jung2006, jung2007,
karrasch2015, steinigeweg2014-2, steinigeweg2015-2}, break integrability of
the XXZ chain and therefore add another layer of complexity. In this context,
improving numerical approaches is imperative to progress in understanding. Within
the class of relevant perturbations, disorder plays a remarkable role since it
goes along with MBL as a new dynamical state of matter. Early on, a numerical work
based on Lanczos diagonalization \cite{karahalios2009} found that, at $\Delta=1$ and
$W=1$, the low-$\omega$ optical conductivity at high $T$ follows the power law
$\text{Re} \, \sigma(\omega) \approx \sigma_\text{dc} + a \, |\omega|^\alpha$,
with $\alpha \approx 1$, being different from Mott's law for the Anderson insulator
$\alpha\approx 2$. Such $\alpha$ was also observed for small but finite $\Delta$
and in a wider range of $W$ \cite{barisic2010}. A more recent theoretical study
\cite{gopalakrishnan2015}  has suggested that $\alpha\rightarrow1$  when approaching
the MBL transition from the  localized ($\sigma_\text{dc}=0$) side, attributed to
rare metallic regions, in contrast to $\alpha \approx 2$, due to rare resonant pairs
deep in the localized phase.

In this paper, we study the optical conductivity in disordered systems using
complementary numerical methods, with a particular focus on dynamical quantum
typicality (DQT) \cite{elsayed2013, steinigeweg2014-1, steinigeweg2015-1} (see
also \cite{gemmer2003, goldstein2006, popescu2006, reimann2007, white2009,
bartsch2009, bartsch2011, sugiura2012, sugiura2013, hams2000, iitaka2003,
iitaka2004}). This method employs the fact a single pure state can exhibit
properties identical to that of the complete statistical ensemble. This fact
has been demonstrated in nontrivial phases of the disorder-free XXZ chain and
allows to study the long-time dynamics of quantum systems with Hilbert spaces
being much larger than those accessible to standard ED approaches. While in
localized phases it is clear that a single {\it eigenstate} cannot be a typical
representative, i.e., the eigenstate thermalization hypothesis (ETH)
\cite{deutsch1991, srednicki1994, rigol2008} is not satisfied, we show
for finite systems that DQT, which is {\it different} from ETH, works
well, i.e., still the overwhelming majority of states drawn at random from a
high-dimensional Hilbert space are typical. 

To outline, we apply DQT to disordered XXZ chains  and demonstrate that a single
pure state can indeed represent the full statistical ensemble within the entire
range from the thermal to the MBL phase. In particular, we find that $\text{Re} \,
\sigma(\omega) \approx \sigma_\text{dc} + a \, |\omega|^\alpha$ with $\alpha
\approx 1$  in a wide range of parameters within the thermal phase and close to
the transition. Moreover, we detail the dependence of $\sigma_\text{dc}$ on $W$
and connect to known results on either very small or very large $W$.  Finally,
we determine the $T$ dependence of $\sigma_\text{dc}$ down to low $T$ in the
thermal phase. We find that this dependence is consistent with the existence
of an MBL mobility edge. Thus, our results provide for a comprehensive picture
of dynamical phases in disordered XXZ chains, as illustrated in Fig. \ref{Fig1}.

{\it Model.} We study the antiferromagnetic XXZ spin-$1/2$
chain with periodic boundary conditions, given by ($\hbar = 1$)
\begin{equation}
H = J \sum_{r=1}^L (S_r^x S_{r+1}^x + S_r^y S_{r+1}^y + \Delta \,
S_r^z S_{r+1}^z + B_r \, S_r^z ) \, , \label{hamiltonian}
\end{equation}
where $S_r^{x,y,z}$ are the components of spin-$1/2$ operators at site $r$.
$J > 0$ is the exchange coupling constant, $L$ the total number of sites,
and $\Delta$ the anisotropy. The local magnetic fields $B_r$ are drawn at
random from a uniform distribution in the interval $[-W,W]$. Thus,
translation invariance and integrability of the model are broken for any $W \neq
0$. Total magnetization $S^z$ is strictly conserved for any value of $W$.
This model has been studied extensively in the context of MBL at $\Delta=1$ and
several exact-diagonalization studies find an MBL phase at infinite temperatures
for $W/J \gtrsim 3.5$ \cite{pal2010, luitz2015}. In this paper, we study the
grand-canonical ensemble $\langle S^z \rangle = 0$, taking into account all
$S^z$ sectors.

\begin{figure}[t]
\centering
\includegraphics[width=0.90\columnwidth]{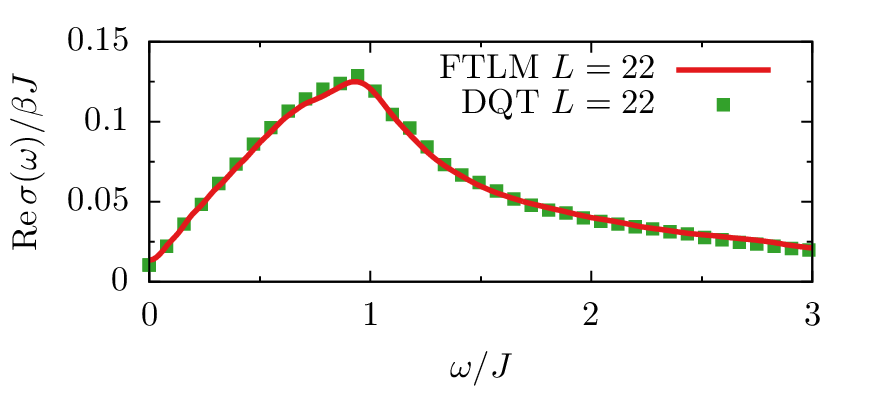}
\caption{(Color online) Comparison of DQT ($t \, J \leq 40$) and FTLM ($M=400$):
$\text{Re} \, \sigma(\omega)$ at $\beta \to 0$, $\Delta = 1$, and $W/J = 2$ for 
$L=22$ and $N=200$. The excellent agreement clearly shows the validity of
typicality. Such agreement is also found for other values of $W$, see \cite{SM}.}
\label{Fig2}
\end{figure}

\begin{figure*}[t]
\centering
\includegraphics[width=1.95\columnwidth]{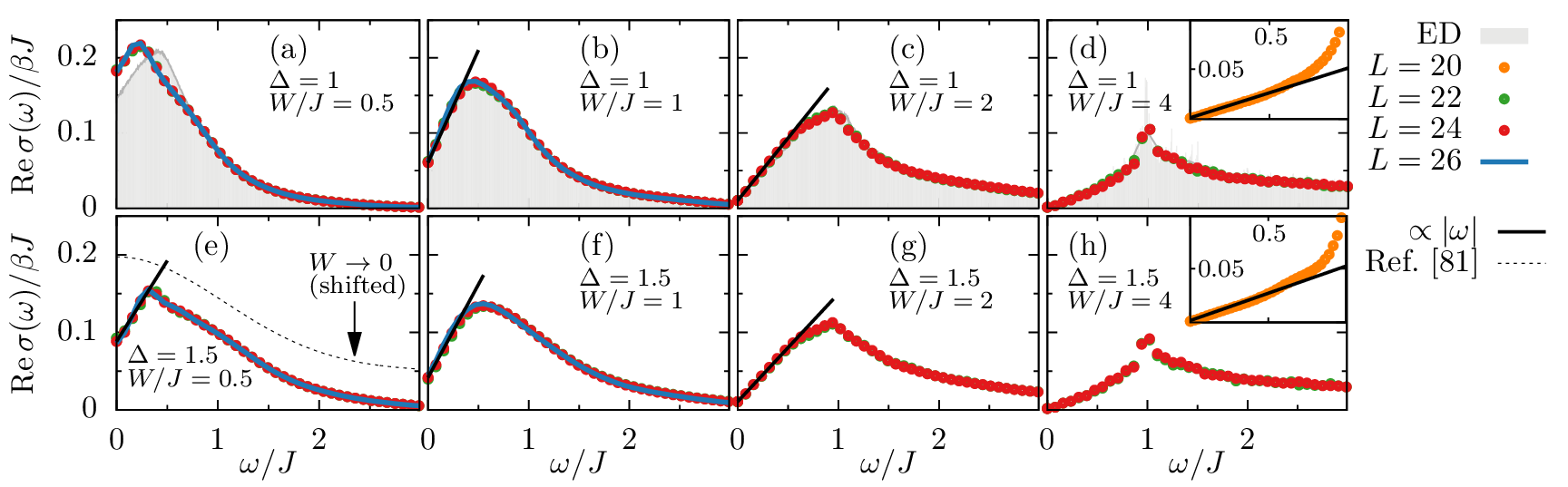}
\caption{(Color online) $\text{Re} \, \sigma(\omega)$
at $\beta \to 0$ for $\Delta = 1.0$ (upper row) and $\Delta = 1.5$ (lower row)
in the transition from small $W / J = 0.5$ (l.h.s.) to strong $W /J = 4$ (r.h.s.)
for the ensemble $\langle S^z \rangle = 0$, as obtained numerically for $L= 14$
using ED and $L > 14$ using DQT ($t \, J \leq 40$; $L < 26$: $N=200$, $L=26$:
$N=20$). For $W = 4$, $L= 20$ data are shown for $N = 10000$ and $t \, J \leq
120$ (insets), reducing statistical errors and increasing frequency resolution.
In all cases (a)-(h), the low-$\omega$ behavior is well described by $\text{Re}
\, \sigma(\omega) \approx \sigma_\text{dc} + a \, |\omega|$ (lines). In (e) the
perturbative result of \cite{steinigeweg2011} for $W \to 0$ is depicted
\cite{SM}.}
\label{Fig3}
\end{figure*}

The spin-current operator $j = J \sum_r (S_r^x S_{r+1}^y - S_r^y
S_{r+1}^x)$ follows from the continuity equation. We are interested in
the autocorrelation function at inverse temperatures $\beta = 1/T$
($k_B = 1$), $C(t) = \text{Re} \, \langle j(t) \, j \rangle/L$,
where the time argument of $j$ has to be understood w.r.t.\ the Heisenberg
picture, $j = j(0)$, and $C(0) = J^2/8$ at high temperatures $\beta \to
0$. From $C(t)$, we determine the optical conductivity via the
Fourier transform
\begin{equation}
\text{Re} \, \sigma(\omega) = \frac{1-e^{-\beta \omega}}{\omega}
\int_0^{t_\text{max}} \text{d}t \, e^{i\omega t}C(t) \, ,
\end{equation}
where the cut-off time $t_\text{max}$ has to be chosen much larger
than the relaxation time $\tau$, with $C(\tau)/C(0) = 1/e$
\cite{steinigeweg2014-2, steinigeweg2015-2}. Note that, using
the Jordan-Wigner transformation, $H$ can be mapped onto
interacting spinless fermions. In this picture, $B_r$ is a
discorded on-site chemical potential and $j$ is the particle current.

{\it Methods.} We use the DQT method, which is most conveniently formulated
in the time domain $t$ and relies on the relation
\begin{equation}
C(t) = \text{Re} \frac{\langle \Phi_\beta(t) | j | \varphi_\beta(t)
\rangle}{L \, \langle \Phi_\beta(0) | \Phi_\beta(0) \rangle} +
\epsilon \, , \label{DQT}
\end{equation}
where $|\Phi_\beta(t) \rangle = e^{-\imath H t -\beta H/2} \, | \psi
\rangle$, $| \varphi_\beta(t) \rangle = e^{-\imath H t} \, j \,
e^{-\beta H/2} \, |\psi \rangle$, and $|\psi \rangle$ is a {\it
single} pure state drawn at random. Most important, the remainder 
$\epsilon$ scales inversely
with the partition function, i.e., $\epsilon$ is exponentially small
in the number of thermally occupied eigenstates \cite{elsayed2013,
steinigeweg2014-1, steinigeweg2015-1}. The great advantage of Eq.\
(\ref{DQT}) is that it can be evaluated without any diagonalization
by using forward-iterator algorithms. In this paper, we employ a
fourth-order Runge-Kutta iterator with a discrete time step $\delta t
\, J = 0.01 \ll 1$. Using this iterator, together with sparse-matrix
representations of operators, we can reach system sizes as large as
$L=30$. However, since we have to average over $N \gg 1$ disorder
realizations, we consider $L \leq 26$.

To additionally corroborate our DQT results, we employ ED for $L=14$
and the finite-temperature Lanczos method (FTLM), formulated in the
frequency domain $\omega$ and yielding $\text{Re} \, \sigma(\omega)$
with a frequency resolution $\delta \omega \propto 1/M$ \cite{prelovsek2013},
where $M \sim 400$ is the number of Lanczos steps used.

{\it Results.} We now present our DQT results, starting with the
infinite-temperature limit $\beta \to 0$. As long not stated otherwise,
all DQT data are obtained from real-time data $t \, J \leq 40$, where the
autocorrelation function $C(t)$ decays fully to zero \cite{SM}. This
finite-time window yields a frequency resolution $\delta \omega / J
\approx 0.08$. 

First, for medium disorder $W / J = 2$, we compare in Fig.\ \ref{Fig2} the
optical conductivity $\text{Re} \, \sigma(\omega)$, as obtained from DQT
and FTLM for a system of size $L=22$. The excellent agreement clearly
shows that a single pure state, drawn at random from a high-dimensional
Hilbert space, is a typical representative of the full statistical ensemble.
This demonstration of typicality in disordered systems of finite size
constitutes a first central result of our paper and is the fundament
for using DQT as an accurate numerical method, for this and other values
of $W$ \cite{SM}.

In Fig.\ \ref{Fig3} we summarize our optical-conductivity results $\text{Re} \,
\sigma(\omega)$ for $\Delta=1.0$ (upper row) and $\Delta = 1.5$ (lower row)
along the transition from small disorder $W/J = 0.5$ (l.h.s.) to strong
disorder $W/J = 4.$ (r.h.s.). Several comments are in order. First, while
finite-size effects increase as $W$ decreases, we find no significant $L$
dependence for large $L \geq 22$ in the disorder range $0.5 \leq W/J \leq 4.0$,
depicted in Fig.\ \ref{Fig3}. Second, while averaging over disorder realizations
is more important for larger $W$, statistical errors for $N = 200$ are already
smaller than the symbol size used for each $W$ shown. Third, despite the large
difference in $L$, the overall agreement with ED data, depicted for $L=14$ in
Fig.\ \ref{Fig3} (a)-(d), proves again that typicality is remarkably well
satisfied. Finally, it is evident from Fig.\ \ref{Fig3} (a) that already at
high $T$ finite-size effects can be significant for $L = 14$.

\begin{figure}[t]
\centering
\includegraphics[width=0.90\columnwidth]{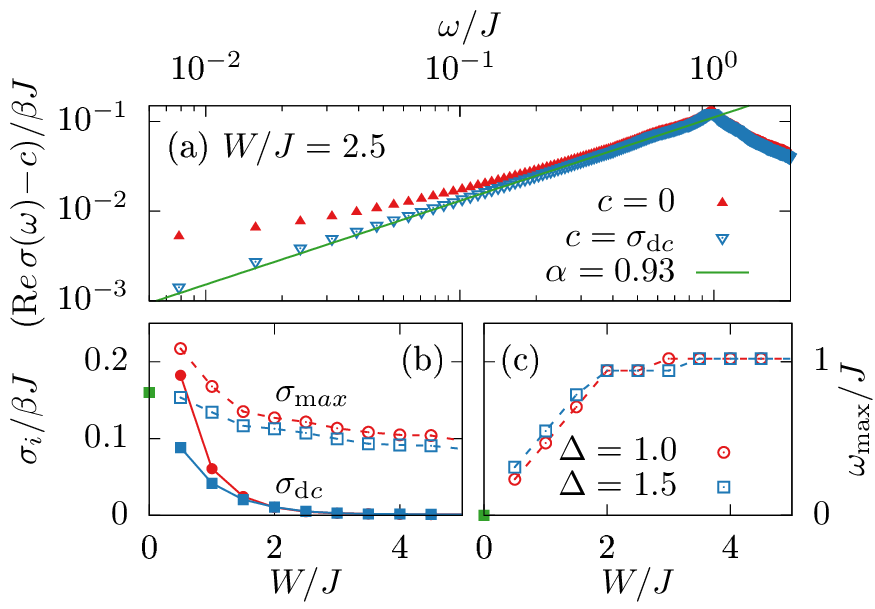}
\caption{(Color online) 
(a) Log-log plot of $\text{Re} \, \sigma(\omega) -c$, with $c = 0$ and
$c = \sigma_\text{dc}$, at $W = 2.5$, $\Delta = 1$, and $\beta \to 0$
($\langle S^z \rangle = 0$, $L=20$, $t \, J \leq 400$, $N = 1000$) as well
as a power-law fit with the exponent $\alpha = 0.93$ being close to $1$. (b),
(c) Disorder dependence of $\sigma_\text{dc}$, the maximum $\sigma_\text{max}$,
and  its position $\omega_\text{max}$ at $\Delta = 1.0$, $1.5$ and $\beta \to
0$  ($\langle S^z \rangle = 0$, $L=24$, $t \, J \leq 40$, $N = 200$). For $W = 0$,
also the $\Delta = 1.5$ result of, e.g., \cite{steinigeweg2011} is indicated
(green square).}
\label{Fig4}
\end{figure}

As shown in Fig.\ \ref{Fig3}, the optical conductivity $\text{Re} \,
\sigma(\omega)$ has a well-defined value $\sigma_\text{dc}$ at
$\omega = 0$ and a maximum $\sigma_\text{max} > \sigma_\text{dc}$
located at $\omega_\text{max} > 0$ for all $W$ depicted. While
$\sigma_\text{dc}$ decreases fast as $W$ increases, $\sigma_\text{max}$
has a much weaker $W$ dependence, see Fig.\ \ref{Fig4} (b). In particular the
position $\omega_\text{max}$ moves to higher frequencies and eventually
saturates at large $W$, see Fig.\ \ref{Fig4} (c). Most notably, for $\omega
\ll \omega_\text{max}$ the optical conductivity is well described by a power
law, i.e., $\text{Re} \, \sigma(\omega) \approx \sigma_\text{dc} + a \,
|\omega|^\alpha$, where $\alpha \approx 1$. The exponent $\alpha = 1$ has
been proposed in \cite{gopalakrishnan2015} at the MBL transition. We
find this exponent also in a wide range of the thermal phase. This finding
does not depend on the frequency resolution and the disorder average, see Fig.\
\ref{Fig3} (d), (h), and can be substantiated by a log-log plot after
subtracting $\sigma_\text{dc}$, see Fig.\ \ref{Fig4} (a). We further
checked that our finding is true for binary disorder \cite{SM}. Note that
the above power-law is different to Mott's law $\text{Re} \, \sigma(\omega)
\propto \omega^\alpha$ with $\alpha = 2$, valid for $W / \Delta \gg 1$
\cite{gopalakrishnan2015}. Moreover, it differs from a subdiffusive
power law with $\sigma_\text{dc} = 0$ and $\alpha < 1$ \cite{agarwal2015,
khait2016}, in agreement with Ref.\ \cite{barisic2016}.

For $W \to 0$, Fig.\ \ref{Fig4} (b), (c) suggests $\omega_\text{max}
\to 0$ and $\sigma_\text{dc} = \sigma_\text{max}$ for $\Delta = 1.0$ and $1.5$. On
the one hand, this suggestion is in line with results at $W = 0$ for
$\Delta = 1.5$ in \cite{prelovsek2004, steinigeweg2011, karrasch2014}. On
the other hand, for $\Delta = 1.0$, the complete form of $\text{Re}\,
\sigma(\omega)$ vs.\ $\omega$ is still under scrutiny \cite{heidrichmeisner2003,
sirker2009, sirker2011, grossjohann2010, herbrych2012}, including the
existence of a finite $\sigma_\text{dc}$.

\begin{figure}[t]
\centering
\includegraphics[width=0.90\columnwidth]{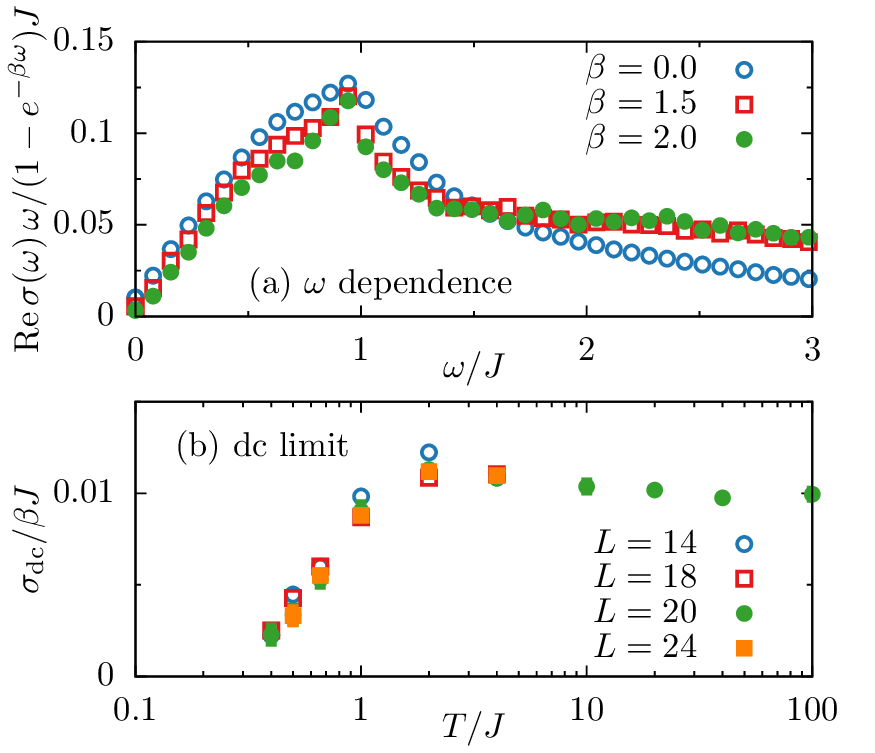}
\caption{(Color online) (a) $\text{Re} \, \sigma(\omega)$ at intermediate $W/J
= 2$ and various $\beta \, J \leq 2$ for $\Delta = 1$ ($\langle S^z \rangle = 0$,
$L=24$, $t \, J \leq 40$, $N = 1000$). (b) Temperature dependence of $\sigma_\text{dc}$
for different $L \leq 24$. (Small error bars for the two largest $L = 20$ and $24$
indicate the difference between $N=500$ and $1000$.) This temperature dependence is
consistent with a mobility edge located at $E- E_\text{min} \sim 2 \, J$.}
\label{Fig5}
\end{figure}

Next, we turn to lower temperatures $\beta \neq 0$, focusing on $\Delta = 1$
and $W = 2$, where $\sigma_\text{dc}$ is already small but still nonzero at
$\beta = 0$. In Fig.\ \ref{Fig5} (a) we depict our results for $\text{Re} \,
\sigma(\omega) \, \omega / (1-e^{-\beta \omega})$, i.e., the mere Fourier
transform of $C(t)$, for various $\beta J \leq 2$ and a single $L=24$.
Clearly, spectral weight at $\omega/J \gtrsim 2$ increases as $\beta$
increases, while the overall structure at $\omega/J \sim 1$ only
weakly depends on $\beta$. In Fig.\ \ref{Fig5} (b) we show the temperature
dependence of $\sigma_\text{dc}$, which is well converged for $L \geq 20$
and $N \geq 500$ in the entire temperature range depicted. Apparently, at
high temperatures, $\sigma_\text{dc} / \beta \approx \text{const.}$ For
$T/J \lesssim 2$, however, $\sigma_\text{dc} / \beta$ decreases rapidly
as $T$ decreases. This finding is a central result of our paper. It
is very suggestive of an interpretation in which extended states are
frozen out below an energy scale of order $E - E_\text{min} \sim 2 J$.
Speaking differently, this result points to the existence of a {\it mobility
edge} in terms of $E$, where $E_\text{min}$ refers to the lower bound
of the spectrum. Similar results have been reported in \cite{karahalios2009}
for smaller values of $W$.

{\it Summary and Conclusion.} We studied the frequency dependence of the
optical conductivity $\text{Re} \, \sigma(\omega)$ of the XXZ spin-$1/2$
chain in the transition from a thermal to a many-body localized phase induced
by the strength of a spatially random magnetic field. To this end, we used
numerical approaches to large system sizes, far beyond the applicability of
standard ED, with a particular focus on DQT. In particular, we showed that the
DQT approach represents a powerful tool to study  dynamical responses of MBL
systems.  First, we demonstrated the validity of typicality in disordered systems. 
Then, we found that the low-frequency behavior of $\text{Re} \, \sigma(\omega)$ 
is well described by $\text{Re} \, \sigma(\omega) \approx \sigma_\text{dc} + a \,
|\omega|^\alpha$, with a constant $\alpha \approx 1$ in a wide range of
the thermal phase and close to the transition. We further detailed the
decrease of $\sigma_\text{dc}$ as a function of increasing disorder or
decreasing temperature. We particularly found that the temperature
dependence is consistent with the existence of a mobility edge. 

{\it Acknowledgments.} We thank X.\ Zotos, P.\ Prelov\v{s}ek, and
F.\ Heidrich-Meisner for fruitful discussions.
J.H.\ acknowledges support by the EU program FP7-REGPOT-2012-2013-1
under Grant No.\ 316165.
Work of W.B.\ and F.P.\ has been supported in part by the DFG through SFB 1143 and
the NSF under Grant No. NSF PHY11-25915. W.B. also acknowledges kind
hospitality of the PSM, Dresden.
R.S.\ thanks the MPIPKS, Dresden and the CCQCN, Crete for
kind hospitality.



\setcounter{figure}{0}
\setcounter{equation}{0}
\renewcommand*{\citenumfont}[1]{S#1}
\renewcommand*{\bibnumfmt}[1]{[S#1]}
\renewcommand{\thefigure}{S\arabic{figure}}
\renewcommand{\theequation}{S\arabic{equation}}

\section{\large Supplemental Material}


\subsection{Time Dependencies}

\subsubsection{Intermediate Times}

\begin{figure*}[t]
\centering
\includegraphics[width=1.95\columnwidth]{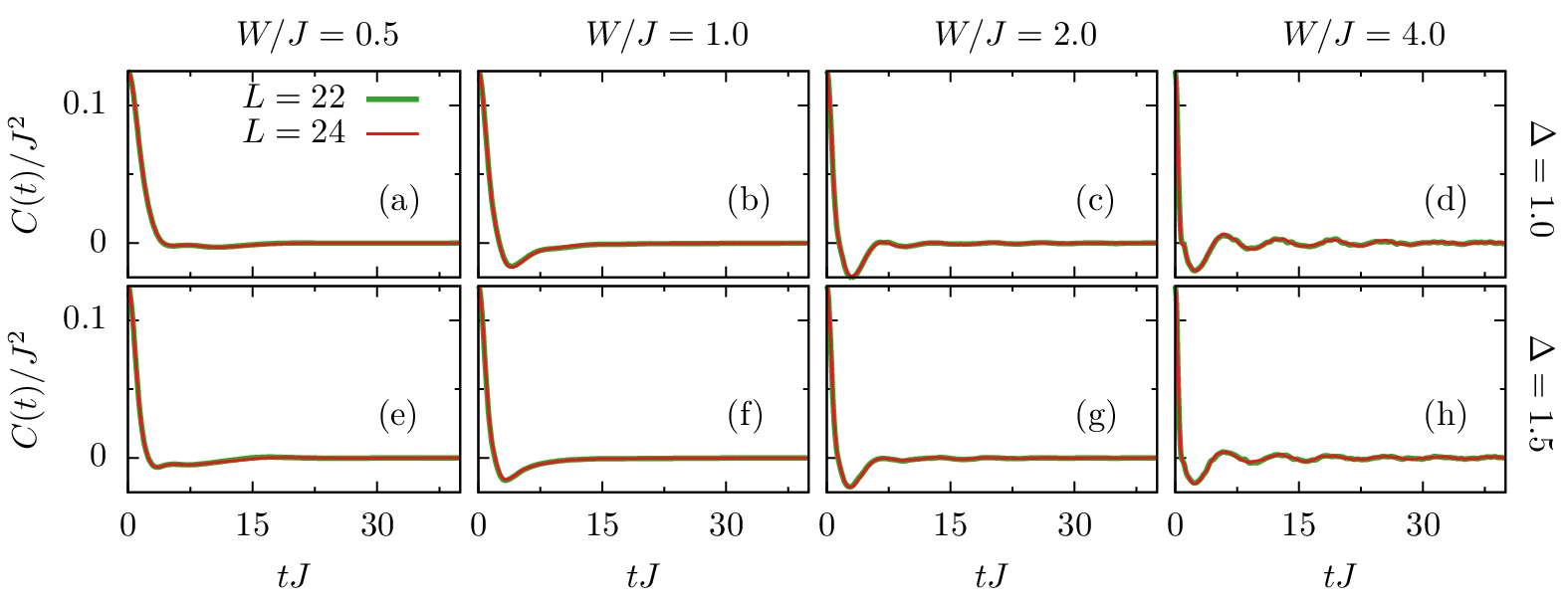}
\caption{(Color online) Data underlying Fig.\ 3 of the main text: Real-time decay of the
high-temperature current autocorrelation function $C(t)$ of the spin-$1/2$ Heisenberg
chain at (a)-(d) $\Delta = 1.0$ and (e)-(h) $\Delta = 1.5$ in the transition from (a),
(e) small disorder $W / J = 0.5$ over intermediate disorder (b), (f) $W /J = 1.0$, (c),
(g) $W / J = 2.0$ to strong disorder (d), (h) $W / J = 4.0$, as obtained numerically
for the grand-canonical ensemble $\langle S^z \rangle = 0$ and system sizes $L = 22$
and $24$. The results shown are averaged over $N = 200$ different disorder realizations
using a uniform distribution $[-W,W]$. The sum rule is $C(0) / J^2 = 0.125$. In all
cases (a)-(h), $C(t)$ decays fully on a time scale $t \, J \leq 40$. Damping of $C(t)$
after its first zero crossing causes the linear $\omega$ dependence $\text{Re} \,
\sigma(\omega) \approx \sigma_\text{dc} + a |\omega|$.}
\label{FigS1}
\end{figure*}

In our paper, we discussed the frequency dependence of the
optical conductivity $\text{Re} \, \sigma(\omega)$ rather than
the time dependence of the spin-current autocorrelation function
$C(t)$ as such. However, we determined $\text{Re} \, \sigma(\omega)$
via the finite-time Fourier transform of $C(t)$. Furthermore, it is
instructive to discuss the real-time decay of $C(t)$. Thus, we
show in Fig.\ \ref{FigS1} the time-dependent data underlying
Fig.\ 3 in our paper.

Clearly, for all $0.5 \leq W/J \leq 4.0$ and the two $\Delta = 1.0$
and $1.5$ depicted, the initial value $C(0)$ agrees well with the
sum rule $J^2/8$, which also shows the accuracy of our DQT approach. 
For all $W$ depicted, the initial decay of $C(t)$ is fast with a
relaxation time $\tau \, J \ll 10$. It is clearly visible that $C(t)$
develops oscillatory behavior for large $W$, which is the origin
of the maximum $\sigma_\text{max}$ located at the position
$\omega_\text{max}$, as discussed in our paper. However, all
oscillations fully decay on a time scale $t \, J \leq 40$ and
there is no signature of a conserved Drude-weight contribution
to $C(t)$ in the long-time limit. Therefore, $t \, J \leq
40$ data is sufficient to precisely determine the $\omega$
dependence of $\text{Re} \, \sigma(\omega)$ in general and the
value of $\sigma_\text{dc}$ in particular.

\subsubsection{Long Times}

\begin{figure}[b]
\centering
\includegraphics[width=0.85\columnwidth]{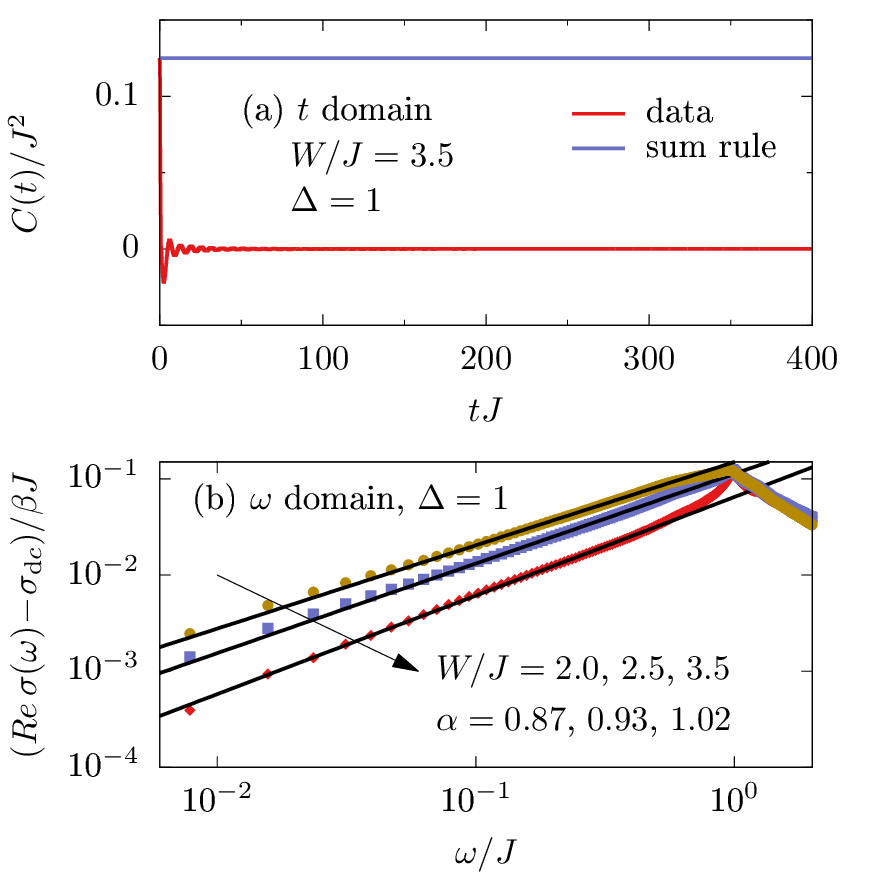}
\caption{(Color online) (a) Real-time decay of $C(t)$ for very long times $t J \leq
400$, averaged over as many as $N=30000$ disorder realizations in a system of size
$L=18$. Remaining parameters: $\Delta = 1$, $W/J=3.5$, and $\beta J \to 0$. (b) Fourier
transform of (a) and, additionally, for $W/J = 2.0, 2.5$ in a log-log plot. Remaining
parameters: identical to (a) except for $N = 5000$. Note that $\sigma_{\mathrm dc}$ is
subtracted from the Fourier transform. Power-law fits are also indicated.}
\label{FigS2}
\end{figure}

In Fig.\ \ref{FigS2} (a) we show $C(t)$ at $\Delta = 1$ and
$W/J = 3.5$ for even longer times $t J \leq 400$ and as many
as $N = 30000$ disorder realizations in a system of size $L=18$.
Clearly, $C(t)$ is practically zero for $t J \gtrsim 50$. Thus, while
taking into account $t J \leq 400$ in the Fourier transform certainly
increases frequency resolution, we find no change of the linear
frequency dependence down to a rather small scale of frequency, see
Fig.\ \ref{FigS2} (b). It is worth mentioning that the $W/J = 3.5$
calculation depicted in Fig.\ \ref{FigS2} took about $20$ CPU years in
total.

In Fig.\ \ref{FigS2} we also depict high-resolution data for the
same set of parameters except for $W/J = 2.0$, $2.5$ and $N=5000$, cf.\
Fig.\ 4 in the main text. Apparently, this data is well described by power
laws with the exponent $\alpha$ being close to 1 in both cases.

\subsection{Binary Disorder}

\begin{figure}[b]
\centering
\includegraphics[width=0.85\columnwidth]{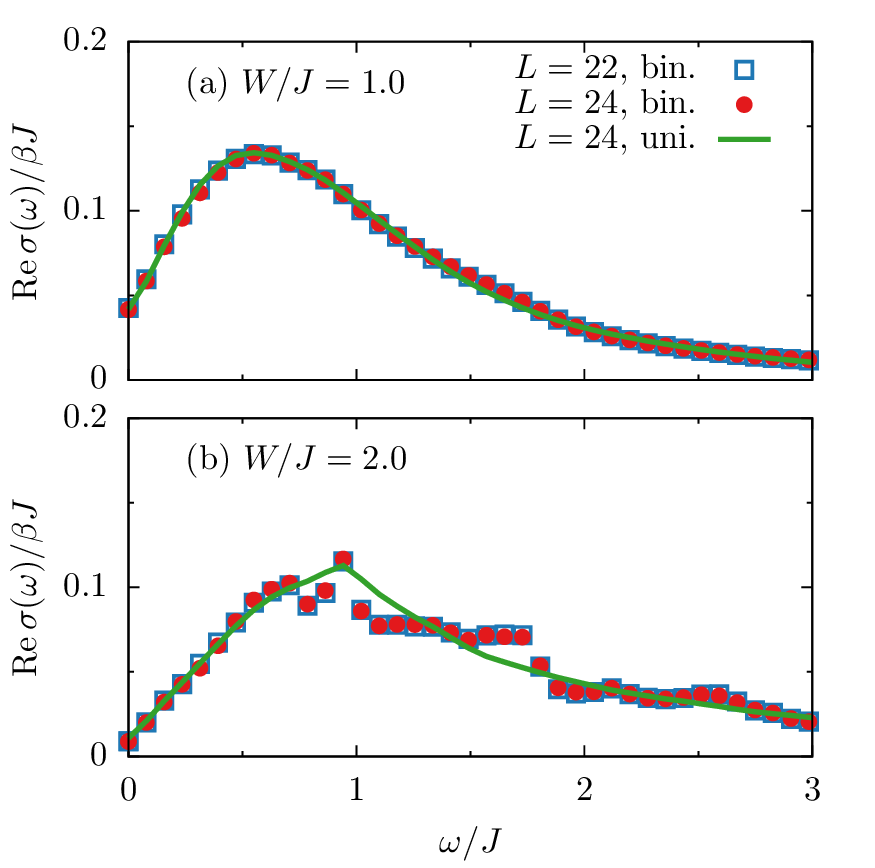}
\caption{(Color online) Optical conductivity $\text{Re} \, \sigma(\omega)$
for binary and uniform distribution and disorder strength (a)
$W/J = 1$ and (b) $W/J = 2$. Remaining parameters: $\Delta=1.5$,
$\beta \, J \to 0$, $L \leq 24$, $t \, J \leq 40$, and $N=200$.
Clearly, the low-$\omega$ behavior does not depend on the specific
probability distribution used, while high-$\omega$ differences
emerge for large $W$.}
\label{FigS3}
\end{figure}

Our paper focused on local magnetic fields $B_r$ drawn at
random from a uniform distribution in the interval $[-W,W]$. To
demonstrate that the results presented do not depend on the
specific distribution used, we repeat the $\Delta=1.5$ calculations
for $W/J = 1.0$ and $W/J = 2.0$ in Fig.\ 3 (f) and (g) for
a binary distribution with the same width, i.e., $B_r = \pm \sqrt{3}
\, W$. In Fig.\ \ref{FigS3} we compare the corresponding results.
Evidently, the low-$\omega$ behavior of the optical conductivity
$\text{Re} \, \sigma(\omega)$  is the same for both distributions,
while differences can only be seen in the high-$\omega$ behavior,
emerging for strong disorder $W$. These differences are not relevant
for the physics discussed in our paper.

\subsection{Finite-Size Effects:\\ Comparison to Clean Systems}

\begin{figure}[t]
\centering
\includegraphics[width=0.85\columnwidth]{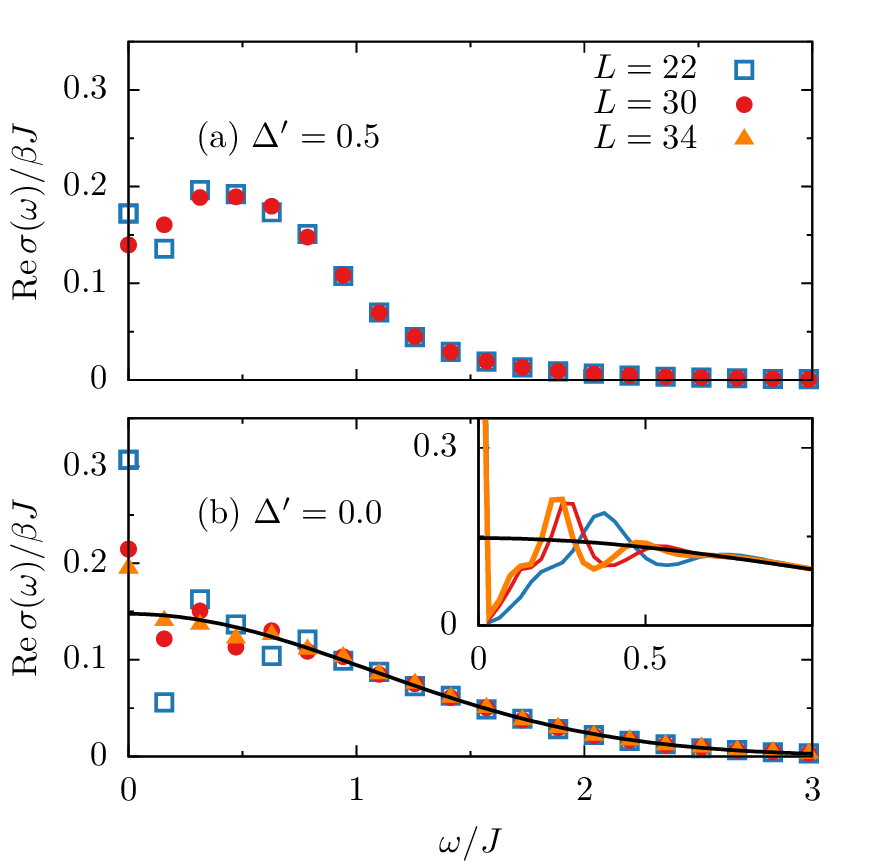}
\caption{(Color online) Optical conductivity $\text{Re} \, \sigma(\omega)$ for
cases without disorder but with (a) $\Delta' = 0.5$ (non-integrable) and (b)
$\Delta' = 0.0$ (integrable). Remaining parameters: $\Delta=1.5$, $\beta
\, J \to 0$, $L \leq 34$, $t \, J \leq 20$, and $N=1$. The solid line in
(b) is the perturbative result of \cite{Ssteinigeweg2011}.
Inset: (a) for $t \, J \leq 100$, i.e., higher $\omega$ resolution.}
\label{FigS4}
\end{figure}

We found in our paper that the optical conductivity $\text{Re} \,
\sigma(\omega)$ has a maximum $\sigma_\text{max} > \sigma_\text{dc}$
located at a position $\omega_\text{max} > 0$. Moreover, we observed
little finite-size effects for large $L \geq 22$. Even though not
expected, we cannot exclude a very slow convergence to the thermodynamic
limit $L \to \infty$. Note that estimating potential finite-size effects
on the basis of the non-interacting case $\Delta = 0$ is not meaningful
for two reasons: First, for $\Delta = 0$ and $W > 0$, the localization
length represents a natural scale for finite-size effects but is absent
in the thermal phase of the $\Delta > 0$ problem. Second, also the case
$\Delta = W = 0$ is well-known to feature huge finite-size effects
because of the highly degenerated spectrum. Moreover, the mean free
path is infinitely large.

\subsubsection{Non-Integrable Systems}

To provide further evidence for finite-size effects being negligibly small,
we compare to results for cases without disorder, i.e., $W = 0$. For such
cases, and $\Delta = 1.5$, the diffusion constant can be estimated
perturbatively along the lines of \cite{Ssteinigeweg2011},
yielding $D/J \sim 0.6$. This value of $D$ corresponds to a mean free path
of a few lattice sites, i.e., the mean free path is small compared to
typical system sizes considered. Thus, finite-size effects are most likely
related to the Hilbert-space dimension being finite and not to a physical
length scale as such. Note that this line of reasoning is also meaningful
for disordered but thermal cases.

We again break the integrability of the XXZ spin-$1/2$ chain but now by
adding to Eq.\ (1), where $W=0$, the next-to-nearest neighbor interaction
\begin{equation}
H' = J \, \Delta' \sum_{r=1}^L S_r^z S_{r+2}^z \label{NNN}
\end{equation}
with the anisotropy $\Delta'$. Adding Eq.\ (\ref{NNN}) does not break
translation invariance and does not change the form of the spin-current
operator.

In Fig.\ \ref{FigS4} (a) we depict the high-temperature optical conductivity
$\text{Re} \, \sigma(\omega)$ for $\Delta = 1.5$ and $\Delta' = 0.5$, for
a large $L = 30$ and a small enough $L = 22$ to illustrate the role of
finite-size effects. It is clearly visible that, as long as $L \ll 30$,
$\sigma_\text{dc}$ decreases with $L$. Hence, together with the overall
convergence at frequencies $\omega / J \gtrsim 0.4$, Fig.\ \ref{FigS4}
(a) proves $\sigma_\text{dc} < \sigma_\text{max}$ in another model. Note
that the largest-subspace dimension for $L = 30$ is comparable to the one
of the $L=26$ disordered model.

This $\omega$ dependence of $\text{Re} \, \sigma(\omega)$ has been found
also in \cite{Smierzejewski2011} using Lanczos diagonalization
and, for spin-$1/2$ ladders, in \cite{Skarrasch2015} using
time-dependent density-matrix renormalization group.

\subsubsection*{Integrable Systems}

Eventually, we contrast all our results presented so far against the
large finite-size effects in the integrable cases $W = \Delta' = 0$,
as shown in Fig.\ \ref{FigS4} (b) for $\Delta = 1.5$. Here, $\text{Re}
\, \sigma(\omega)$ is governed by finite-size effects at $\omega = 0$
and $\omega > 0$. Furthermore, these finite-size effects depend on
the $\omega$ resolution used, see the inset of Fig.\ \ref{FigS4} (b).
Thus, a very careful analysis is needed to determine correctly the
thermodynamic limit \cite{Sprelovsek2004, Ssteinigeweg2011, Skarrasch2014}, 
yielding the dc value $\sigma_\text{dc} / \beta \, J\sim 0.15$. While
this value is depicted in Fig.\ 4 (a) of our paper, none of
our actual results rely on any kind of extrapolation.

\begin{figure}[t]
\centering
\includegraphics[width=0.85\columnwidth]{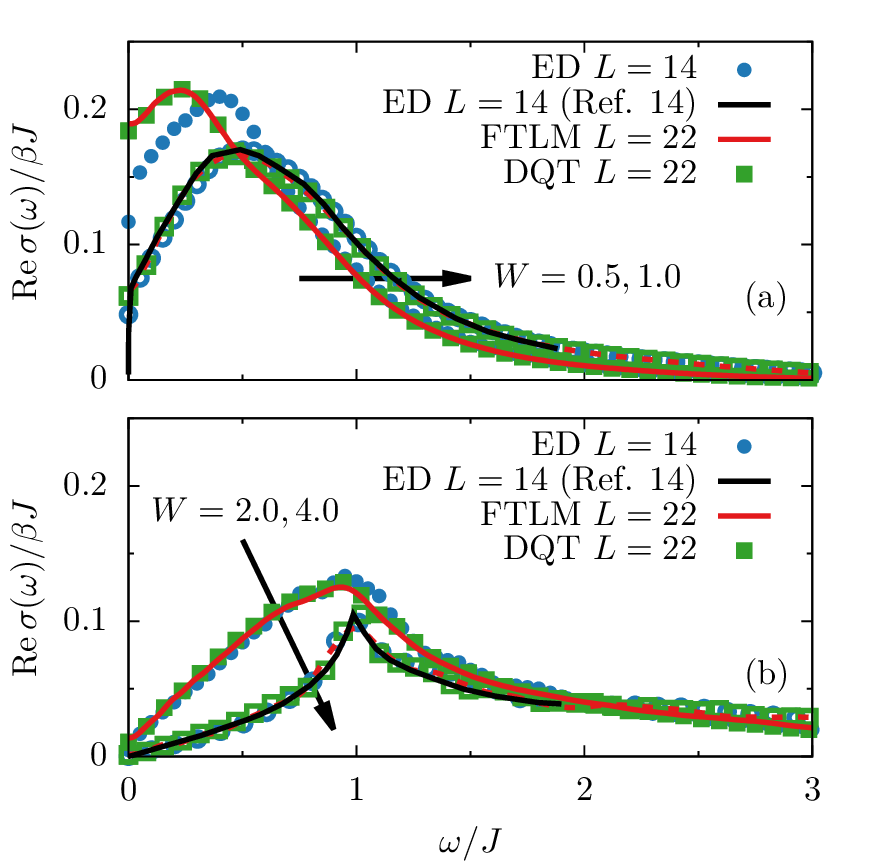}
\caption{(Color online) \label{FigS5} Optical conductivity $\text{Re} \,
\sigma(\omega)$ at high temperatures $\beta \to 0$, as calculated by ED
($L=14$), FTLM ($L=22$), and DQT ($L=22$), for the isotropic point $\Delta
=1.0$ and various disorder strengths (a) $W/J = 0.5$ ,$1.0$ and (b) $W/J =
2.0$, $4.0$. ED data for $W/J = 1.0$, $4.0$ is taken from
\cite{Sgopalakrishnan2015}.}
\end{figure}

\subsection{Comparison to Exact and Lanczos Diagonalization}

To additionally confirm the DQT results presented in our paper, we present
results from two other numerical techniques: standard exact diagonalization
(ED) and the finite-temperature Lanczos method (FTLM) \cite{Sprelovsek2013}.
Both numerical techniques have direct access to the frequency domain. While
ED data is binned in channels of width $\delta \omega /J = 0.005$,
the resolution of FTLM depends on the number of Lanczos steps $M$ and
the energy span $\Delta E$, i.e., $\delta\omega \propto \Delta E/M$.
Here, we use $M=400$. Furthermore, we use $10$ initial random vectors for
each of the $N=100$ disorder realizations, to decease any remaining
statistical error associated with the initial state. 

In Fig.\ \ref{FigS5} we show ED ($L=14$) and FTLM ($L=22$) data for
$\Delta = 1.0$, together with the DQT data ($L=22$) presented in the
main text. The overall agreement of all methods for $W\ge1$ is
remarkably good, despite the smaller system size $L=14$ accessible to ED.
For small $W$, significant finite-size effects are visible for $L=14$.
Note that ED data for $W/J = 1.0, 4.0$ is taken from \cite{Sgopalakrishnan2015}.


\end{document}